\documentclass[aps,prl,twocolumn,superscriptaddress]{revtex4}

\usepackage[T1]{fontenc}
\usepackage[latin9]{inputenc}
\usepackage[english]{babel} 

\usepackage{amsmath}
\usepackage{verbatim}
\usepackage{graphicx}
\usepackage{color}
\usepackage{xcolor}
\usepackage{setspace}
\usepackage{braket}
\usepackage{bbold}
\usepackage{hyperref}

\usepackage{tipa}
\usepackage{amssymb}
\usepackage{wasysym}
\usepackage{bbold}
\usepackage[abs]{overpic}

\usepackage{cancel}

\definecolor{redred}{HTML}{D53E4F}

\newcommand{\Id}{{\mathbb 1}}

\newcommand{\len}{\text{L}}

\begin{document}

\title{Tensor Network Simulation of compact one-dimensional lattice Quantum Chromodynamics at finite density}

\author{Pietro Silvi}
\affiliation{Center for Quantum Physics, and Institute for Experimental Physics, University of Innsbruck and Institute for Quantum Optics and Quantum Information, Austrian Academy of Sciences, A-6020 Innsbruck, Austria}

\author{Yannick Sauer}
\affiliation{Institute for Complex Quantum Systems \& Center for Integrated Quantum Science and Technologies, Universit\"at Ulm, D-89069 Ulm, Germany}

\author{Ferdinand Tschirsich}
\affiliation{Institute for Complex Quantum Systems \& Center for Integrated Quantum Science and Technologies, Universit\"at Ulm, D-89069 Ulm, Germany}


\author{Simone Montangero}
\affiliation{Institute for Complex Quantum Systems \& Center for Integrated Quantum Science and Technologies, Universit\"at Ulm, D-89069 Ulm, Germany}
\affiliation{Theoretische Physik, Universit\"at des Saarlandes, D-66123 Saarbr\"ucken, Germany.}
\affiliation{Dipartimento di Fisica e Astronomia, Universit\'a degli Studi di Padova, I-35131 Italy}


\date{\today}

\begin{abstract}

We perform a zero temperature analysis of a non-Abelian lattice gauge model corresponding to an SU(3) Yang Mills theory in 1+1D at low energies. Specifically, we characterize the model ground states via gauge-invariant Matrix Product States, identifying its phase diagram at finite density as a function of the matter-gauge interaction coupling, the quark filling, and their bare mass.
Overall, we observe an extreme robustness of baryons: For positive free-field energy couplings, all  detected phases exhibit colorless quasiparticles, a strong numerical hint that QCD does not deconfine in 1D.
Additionally, we show that having access to finite-density properties, it is possible to study the stability of composite particles, including multi-baryon bound states, such as the deuteron.

\end{abstract}

\maketitle

The theory ruling the strong interactions within and among hadrons, Quantum ChromoDynamics (QCD), is the main focus of much of the current experimental and theoretical effort in high energy physics~\cite{OpenProblemsQCD}. Many of the collective phenomena arising from this theory, including the phase diagram, have yet to be fully characterized, especially at finite density~\cite{QCDPhasedia98,QCDphasedia06}. In fact, non-perturbative numerical approaches based on lattice formulations~\cite{Wilson74,Kogut1975,Schwingermodel,Hamer1997}, such as MonteCarlo simulations, suffer by the notorious sign problem for complex actions, e.g., in the presence of fermions, such as quarks, or at finite chemical potentials~\cite{Troyer2005}. A promising, alternative route to simulate gauge theories modeled on a lattice based on Tensor Network (TN) ansatz states has been recently put forward.  Indeed, TNs have already shown significant capabilities, delivering quantitative predictions for lattice gauge theories in one spatial dimension \cite{Byrnes2002,LGTN,Tagliacozzo2014,MCBReview2018}. Mostly, Abelian lattice gauge theories have been investigated \cite{Tagliacozzo2011,Rico2014a,Banuls:2013qf,KuhnSaito2,Verstraete:LGT,SchwingerMassMCB2013}, with a few non-Abelian exceptions \cite{PsiSU2,Kuhn2015,SU2MCB2017}. Yet, to our knowledge, no further attempt at capturing the phase properties of a microscopic dynamics analogous to QCD has been made.

Here we present the TN study of a one-dimensional lattice gauge theory with SU(3) symmetry in the quantum link formulation~\cite{ChandraJens,Chandrasekharan1997,Zohar2015}. The model involves flavorless Kogut-Susskind matter fermions~\cite{Kogut1975} and SU(3) Yang-Mills gauge fields, truncated to the first nontrivial Casimir operator excitation.
We investigate the equilibrium properties of this QCD-like model at finite density, at zero temperature and finite lattice spacing $\ell$. We show that TN enables investigation of the confinement problem in models of increasing complexity and  by means of different tools, identifying phases, order properties and binding energies of composite particles.
In particular, for zero quark bare mass, we detect a Luttinger liquid of baryons at finite nonzero charge density.
Only at no charge imbalance the theory does exhibit insulating phases, one spontaneously breaking chiral symmetry 
at low interactions, and another forming dimers in OBC for large interactions. 
Finite bare masses open energy gaps for weak interactions, resulting in insulating phases even at finite density. 
Finally, we show directly evaluating the binding energy that it is more energetically 
favourable to form colorless baryons then colorful particles, and that in this theory the deuteron is not 
energetically stable.

These results foster further development of analogous modelizations in two or higher dimensions, as well as the development of quantum simulation strategies  \cite{Jordan2012,Banerjee2013,tagliacozzoAbel2013,Tagliacozzo2013,ZoharFarax,Martinez:2016kq}, on atomic-molecular-optical platforms, capable of implementing such interesting lattice gauge physics, such as the recently proposed hybrid simulators \cite{KlcoSchwinger2018,KokailSchwinger2018,LuSchwinger2018}.

\begin{figure}[h]
 \includegraphics[width=\columnwidth]{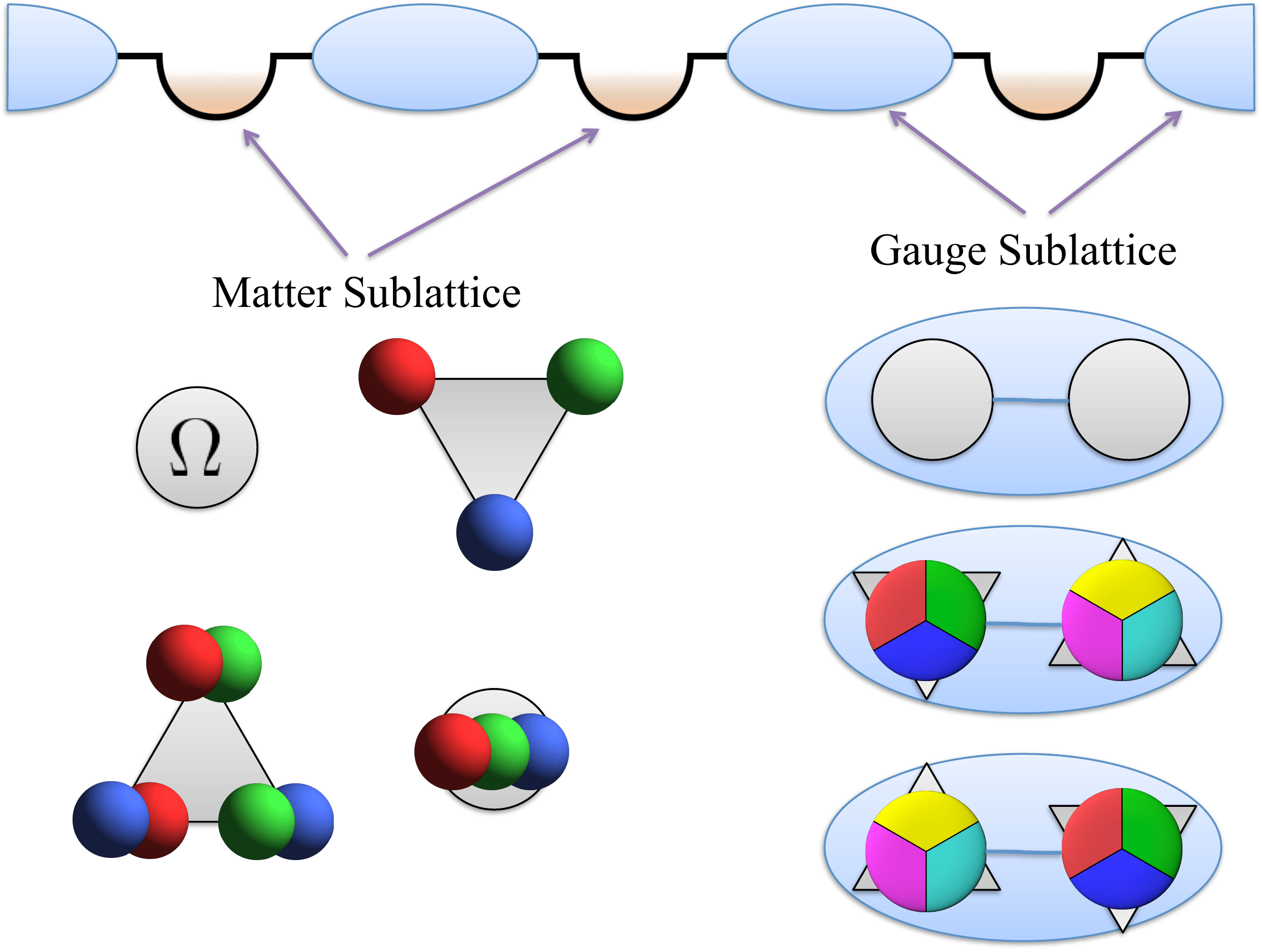}\\
 \begin{overpic}[width = \columnwidth, unit=1pt]{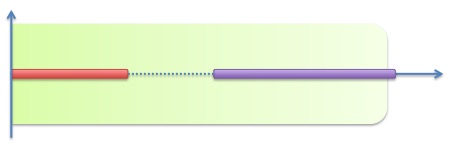}
 \put(0, 267){(a)}
 \put(0, 80){(b)}
 \put(70, 15){Baryon liquid}
 \put(8, 42){Chiral insulator}
 \put(134, 42){Dimer insulator}
 \put(238, 42){$t$}
 \put(10, 68){$\nu$}
 \put(-1, 36){$\frac{3}{2}$}
 \end{overpic}
\caption{ \label{fig:pictorial}
(a) Schematic description of the degrees of freedom. KS fermions live on the matter sublattice: They represent quarks (odd sites) or antiquark-holes (even sites) and have three colors ($r,g,b$).
Listing the 8 local matter states according to the total filling highlights the corresponding $SU(3)$ irrep: The empty state is a color-singlet (grey bullet),
the singly-filled states form a $q$ irrep ($\nabla$ triangle), the doubly-filled states form a $\bar{q}$ irrep ($\Delta$ triangle), and the triply-filled state another color-singlet.
The truncated gluon field space we consider includes 19 states: one singlet-singlet state, nine $q \otimes \bar{q}$ states, and nine $\bar{q} \otimes q$ states.
(b) The phase diagram of the bare-massless case $m=0$, plotted as a function of the filling $\nu$ and the matter-gauge interaction $t$.
}
\end{figure}
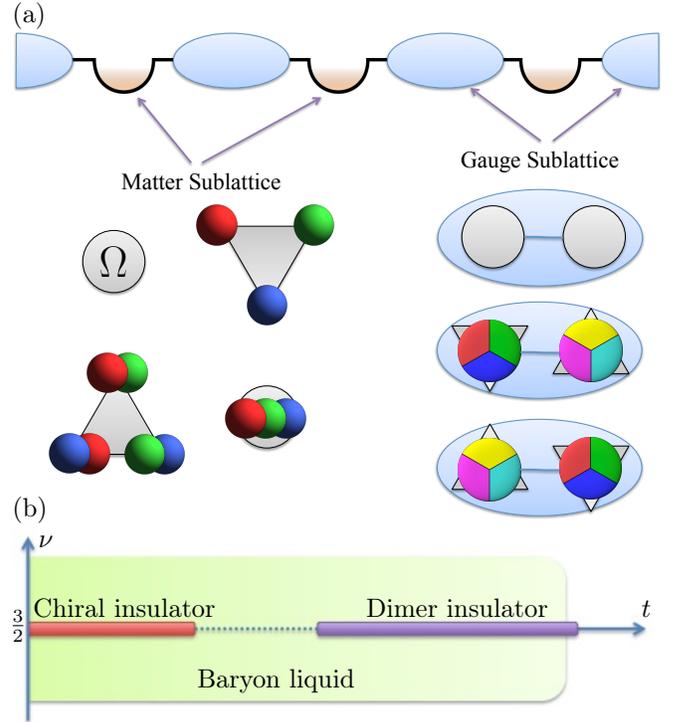

{\it Model $-$} We consider a lattice gauge model in one spatial dimension equipped with a SU(3) local gauge symmetry.
The matter field sublattice, labeled with sites $\{ j \}$, hosts colorful (spinless, flavorless) Kogut-Susskind (KS) fermions \cite{Kogut1975,Kogut1979}:
Odd sites $\{ 2j-1 \}$ host quark fermions, each possessing a fundamental representation $q = (1,0) = \{r,g,b\}$ for its color degree of freedom.
Even sites $\{ 2j \}$ host antiquark fermions, with an antifundamental representation $\bar{q} = (0,1)$. Via a particle-hole transformation performed on even sites, common to KS Hamiltonians, we convert them to be quark-like ($q$) fermions again, as sketched in Fig.~\ref{fig:pictorial}(a).

The gauge field (gluon) sublattice, labeled with links $\{ j,j+1 \}$, is defined via its left-hand $L_{j,j+1}^{\nu}$ and right-hand $R^{\nu}_{j,j+1}$ non-Abelian `electric' field operators
(here $\nu \in \{1,8\}$ corresponds to the 8 generators of the SU(3) group),
and by its parallel transporters operators $U_{j,j+1}^{a,a'}$, with $a,b \in \{r,g,b\}$,  which must obey the commutation relations:
$[L_{j,j+1}^{\nu},U_{j',j'+1}^{a,a'}] = - \frac{1}{2} \delta_{j,j'} \sum_k \lambda^{\nu}_{a,k} U_{j,j+1}^{k,a'}$ and
$[R_{j,j+1}^{\nu},U_{j',j'+1}^{a,a'}] = \frac{1}{2}  \delta_{j,j'}  \sum_k U_{j,j+1}^{a,k} \lambda^{\nu}_{k,a'}$ where the factors 
$ \lambda^{\nu}$ are the 8 Gell-Mann matrices.

The truncated one-dimensional lattice QCD Hamiltonian we adopt commutes with all the Gauss' law generators $Q_{j}^{\nu}$,
detailed in the Supplementary Material (SM),
and reads
\begin{multline} \label{eq:hamzero}
 H_{\text{SU(3)}} = 
 - \frac{t}{\ell} \sum_{j=1}^{\len -1} \sum_{a,a'}^{r,g,b} \left( c^{\dagger}_{j,a} \, U_{j,j+1}^{a,a'} \, c_{j+1,a'} + h.c. \right) 
 \\
 + g \ell \sum_{j=1}^{\len -1} \left( {C^{(L)}_{j,j+1}} + {C^{(R)}_{j,j+1}} \right)
 - m \sum_{j=1}^{\len} \sum_{a}^{r,g,b} (-1)^{j} c^{\dagger}_{j,a} c_{j,a},
\end{multline}
where the Dirac fermionic operators $c_{j,a}^{(\dagger)}$ with $a \in \{r,g,b\}$ describe the KS matter field, and obey the usual Clifford algebra
$\{c_{j,a},c_{j',a'}\} = 0$
and
$\{c_{j,a}^{\dagger},c_{j',a'}\} = \delta_{j,j'} \delta_{a,a'}$.
The operators $C^{(L)}_{j,j+1} = \sum_{\nu=1}^{8} ( L^{\nu}_{j,j+1} )^2$ and  $C^{(R)}_{j,j+1} = \sum_{\nu=1}^{8} ( R^{\nu}_{j,j+1} )^2$
are the quadratic Casimir operators, respectively for the left-hand side and the right-hand side of the gluon at $\{j,j+1\}$, and define the free gluon-field energy term in the Hamiltonian \eqref{eq:hamzero}.
The first line in equation \eqref{eq:hamzero} represents the coupling between matter and gauge fields,
and describe a process of quark-antiquark pair creation/annihilation, which in the staggered fermion language reads as a nearest-neighbor hopping term, while the gluon
in the middle link is updated to protect the Gauss' law.
The last term in the Hamiltonian represents the bare mass of quarks and antiquarks, and it appears as a staggered chemical potential according to the KS prescription. The straight chemical potential is absent as we run simulations at finite quark density.

In this work we consider a static cutoff of the gluon field in accordance to its pure energy density. Precisely we truncate the space right after the first nonzero
eigenvalue of the Casimir $C^{(L)}$ and $C^{(R)}$ (that is $\frac{4}{3}$). Therefore, the gluon space we are considering is composed by
a colorless-colorless $(0,0) \otimes (0,0)$ state with energy $0$, nine quark-antiquark $q \otimes \bar{q}$ states with energy $8g \ell/3$, and
nine antiquark-quark $\bar{q} \otimes q$ states with energy $8g \ell/3$, for a total of 19 gluon field states.
In order to increase the cutoff energy, it may be possible to follow an approach analogous to Ref.~\cite{SU2MCB2017} adapted to SU(3), however, we do not perform this study in the present work.

Hereafter, we fix the energy scale by setting $g = 3/4$ and characterize the ground state $| \Psi_0 \rangle$ of the Hamiltonian \eqref{eq:hamzero}, as a function of
$t$, $m$ and the fermion filling $\nu = \len^{-1} \sum_{j=1}^{\len} \sum_{a}^{r,g,b} c^{\dagger}_{j,a} c_{j,a}$ which is a global symmetry.
Notice that $\nu \in [0,3]$, and that $\nu = 3/2$ represents the KS vacuum sector, where there is no matter-antimatter imbalance.
The aforementioned truncation allows us to implement {\it ab initio} manipulations onto the model, and effectively simplify the
gauge symmetry into a $Z_3$ group. We show explicitly this treatment, carried out via quantum link formulation \cite{ChandraJens,Brower1999},
 in the SM.
In order to numerically simulate the quantum system, we use a Matrix Product State (MPS) description of the many-body quantum state
that embeds both the gauge and the global symmetries \cite{LGTN}, and use a Time-Evolving Block Decimation algorithm in imaginary time to approximate the ground state $| \Psi_0 \rangle$. Running simulations with bondlink dimension of the MPS up to $D \sim 400$ allows us to measure local quantities and correlators with a convergence precision of $\sim 10^{-6}$, sufficient to characterize the phase properties.

\begin{figure}
\includegraphics[width=1.0\columnwidth]{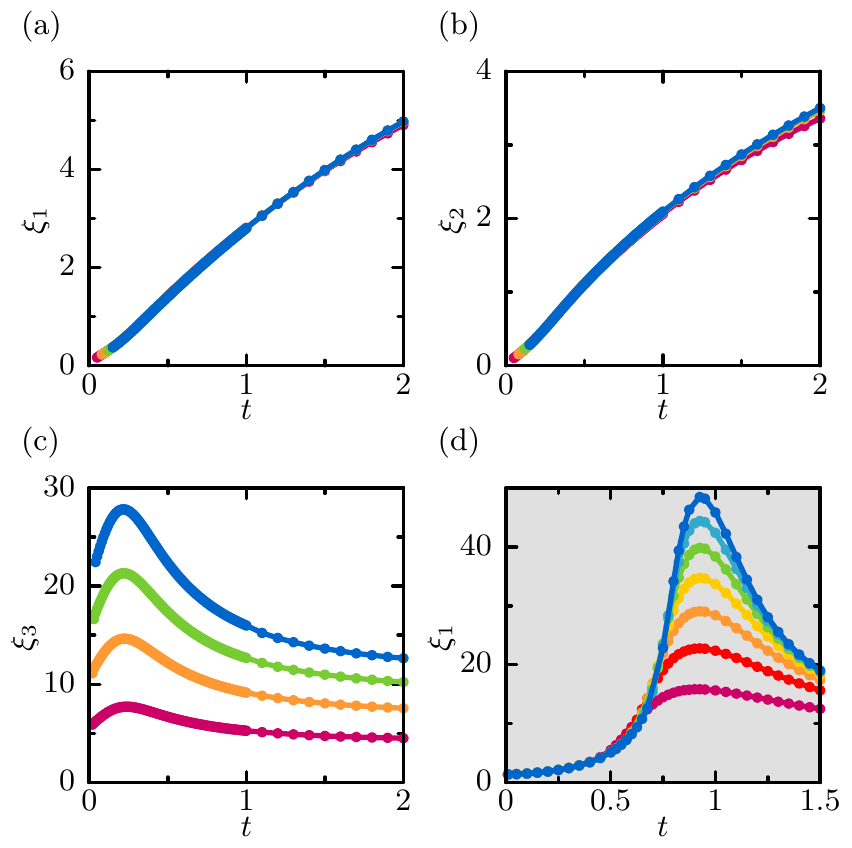}
\caption{ \label{fig:confined}
Correlation lengths  $\xi_\kappa$, with $\kappa = 1,2,3$, of the Luttinger Liquid order parameters at free-field coupling $g>0$ [(a)--(c)] and, for consistency check, $g<0$ (d).
(a--c) Here $\nu = 1$, $m=0$ and $g=1$: only baryonic liquid order emerges.
(d) Here $\nu = -3/2$, $m=0$ and $g=-1$: A quark-liquid appears.
System sizes are $L = 48$ (magenta), $72$ (red), $96$ (orange), $120$ (yellow), $144$ (green), $168$ (cyan), $192$ (blue).
}
\end{figure}

\begin{figure*}
\includegraphics[width=1.0\textwidth]{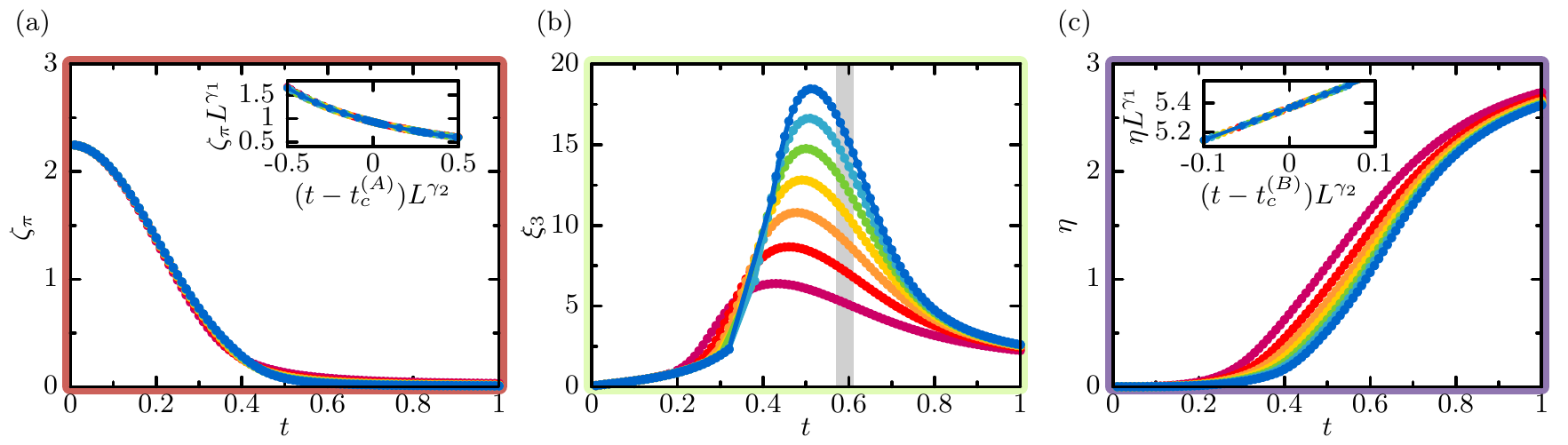}
\caption{ \label{fig:orderpars}
Order parameters at filling $\nu = 3/2$ (KS vacuum), as a function of the matter-gauge interaction $t$: (a) Chiral insulator $\zeta_\pi$, (b) Baryonic liquid $\xi_{3}$ (c) Dimer insulator $\eta$. System sizes are $L = 48$ (magenta), $72$ (red), $96$ (orange), $120$ (yellow), $144$ (green), $168$ (cyan), $192$ (blue).
The shaded region denotes the extrapolated liquid phase $t^{(A)}_{c}<t<t^{(B)}_{c}$, as determined from finite-size scaling analysis (insets). The panel frames are color coded according to the phase diagram of Fig.~\ref{fig:pictorial}, and show the order parameter of the corresponding phase.
}
\end{figure*}

{\it Results: (1) No deconfinement $-$}
We explore various parameter ranges, searching for phases which exhibit Luttinger Liquid (LL) order where the quasiparticles
are respectively quark-like, meson-like, or baryon-like.
Single particle (quark-like) excitations can not be colorless, so they appear with gluon strings attached. It is possible to define a unique string-correlator
acting as order parameter of such quark-like LL behaviour as
$\mathcal{C}_{j,j'} = \sum_{a,a'} c^{\dagger}_{j,a} S^{a,a'}_{j,j'} c_{j',a'}$. The string-body operator $S^{a,a'}_{j,j'}$, which acts on all the gluons between sites $j$ and $j'$,
satisfies important properties: It  transforms covariantly both left-side and right-side, so that $\mathcal{C}_{j,j'}$ is a gauge invariant quantity. Moreover, in order
to capture quasi long-range order
it is a unitary propagator, i.e.,  $\mathcal{C}^3_{j,j'} = (\mathcal{C}_{j,j'})^3  = c^{\dagger}_{j,r} c^{\dagger}_{j,g} c^{\dagger}_{j,b} c_{j',b} c_{j',g} c_{j',r}$.
Two particle (Meson-like) excitations, revealing color BCS order, are captured by $\mathcal{C}^2_{j,j'} = (\mathcal{C}_{j,j'})^2$ and also possess a string attached.
Finally, $\mathcal{C}^3_{j,j'}$ captures a LL order of baryons, or three-particle excitations, which are colorless, and thus carry no string.
For each of these quasi-long range order parameters $\mathcal{C}^{\kappa}_{j,j'}$ with $\kappa \in \{1,2,3\}$, we can evaluate the correlation length $\xi_\kappa$, which is well-approximated by the expression
$\xi_\kappa = \sqrt{  \sum_{l \neq 0} (| l |-1)^2 \bar{\mathcal{C}}^{(\kappa)}_l / \sum_{l \neq 0} \bar{\mathcal{C}}^{(\kappa)}_l }$,
where $\bar{\mathcal{C}}^{(\kappa)}_l = \frac{1}{L-l} \sum_j \langle \Psi_0 | \mathcal{C}^{\kappa}_{j,j+l}  | \Psi_0 \rangle$. Hence, a LL order of $\kappa$-particles
excitations is revealed by a diverging correlation length $\xi_\kappa \to \infty$ for increasing system sizes $\len \to \infty$.
We carefully searched different parametric ranges of the model couplings, and we observed that, as long as $g > 0$ (positive energy densities of the gluon field),
only the baryonic Luttinger liquid $\xi_3$ emerges as a spontaneous quasi long-range order in the truncated lattice QCD model of Eq.~\eqref{eq:hamzero}.
the quark-like and meson-like order parameter exhibit correlation lengths which never go beyond 10 sites (a typical example is shown in Fig.~\ref{fig:confined}.a and \ref{fig:confined}.b)
regardless of $t,m$ and $\nu$, thus such order is not established.
Such observation leads to the conclusion that there are no phases (for $g > 0$) with mobile colored quasiparticles, and thus the SU(3) theory under study is strongly confined.
This is in contrast to the truncated SU(2) Yang-Mills lattice theory previously studied by some of us \cite{PsiSU2}, which instead exhibited liquid phases with 
(colored) single-particle excitations.
For consistency, we remark that such strong confinement effect is energetic and not a byproduct of a symmetry protection. To stress this, we show that simply by setting a (non-physical) negative free-field coupling $g < 0$,
without changing the gluon field truncation, indeed we activate the quark-liquid and meson-liquid orders. This is shown in Fig.~\ref{fig:confined}.d.

\begin{figure}
\includegraphics[width=1.0\columnwidth]{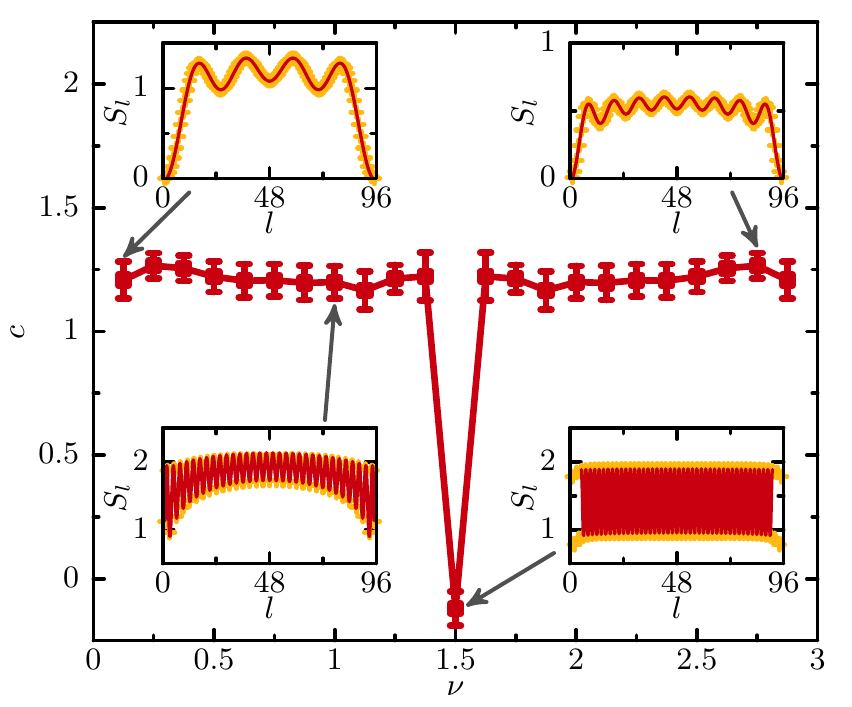}
\caption{ \label{fig:entanglement}
Central charge $c$, as a function of the filling $\nu$ at $t=2.0$, obtained from fitting the theoretical prediction (see text) to the entanglement entropy profile $S_l$, over partition-size 
$l$ at system size $L=96$.
The sharp dip at $\nu = 3/2$ signals the insulating, dimer phase.
The panels show a few examples of the $S_l$ profiles and the corresponding fits.
}
\end{figure}

{\it (2) Phase Diagram $-$}
We explored various parameter regimes, after setting $g = 3/4$, simulating the ground state of the truncated lattice QCD Hamiltonian
for various finite system sizes $\len$. We estimated the phase properties, reported in Fig.~\ref{fig:pictorial}(b), by extrapolating the observable quantities at the thermodynamical limit $\len \to \infty$.
The MPS representation for the variational many-body wavefunction grants access to the entanglement entropies $\mathcal{S}_{l}$ of any left-right
bipartion of the system: $\{1 .. l | l+1 .. \len\}$. We then discriminate between gapped and gapless phases, 
by estimating the central charge $c$ of the correspondent conformal theory (see Fig.~\ref{fig:entanglement} in SM), via
$\mathcal{S}_l \simeq \frac{c}{6} \log\left( \len \sin(\pi l / \len ) \right) + S'_l(k_F) + c' $ \cite{CalabreseCardy}, where
the correction $S'_l(k_F) = a' \cos( 2 k_F (l - \len/2)) \sin^{-b'} (\pi l / \len)$ takes into account Fermi oscillations and allows us to access
the effective fermi wavevector $k_F$, while $a'$, $b'$ and $c'$ are fitting constants \cite{pasqualtweaker1,pasqualtweaker2}.

In the bare massless case ($m = 0$) gapped phases are found only in the KS vacuum, which corresponds to filling $\nu = 3/2$ in the fermion language.
A weak-interacting insulator is detected for $t < t^{(A)}_{c} \simeq 0.57 \pm 0.01$. This phase exhibits chiral order, as it spontaneously breaks the reflection symmetry $P$ centered on a link
(or equivalently the particle-hole symmetry $C$). Such phase appears as a charge-density-wave insulator with wavevector $k = \pi$, and we characterize
it by the order parameter $\zeta_\pi = \sqrt{ \frac{1}{L(L+1)} \sum_{j \neq j'} e^{i \pi (j-j')} \langle (n_j - \nu)(n_j' - \nu) \rangle }$. The latter is reported in Fig.~\ref{fig:orderpars}, where
$n_j = \sum_{a}^{r,g,b} c^{\dagger}_a c_a$
is the local density operator. The presence of this phase at low values of the matter-gluon coupling $t \ll g$ confirms our predictions based
on second-order and third-order degenerate perturbation theories in $t / g$, which are carried out analytically in the SM.
A second insulating phase exhibiting a different type of order is observed at strong matter-gauge interactions $t > t_c^{(B)} = 0.61 \pm 0.01$.
This phase displays a strong staggerization of entanglement, showing higher entanglement when the bipartition is performed on an odd-even bond,
while lower entanglement on an even-odd bond: This behaviour is shown in Fig.~\ref{fig:entanglement} (bottom right Panel). We interpret this effect as
the formation of entangled dimers, resembling what is found in the J1-J2 model~\cite{HaldaneJ1J2dimer}.
If the boundary conditions were periodic, we would expect to see a translationally-invariant resonant valence bond state. However, 
due to the presence of open boundaries, a specific dimer state is selected, with each entangled pair sits on an odd-even $\{2j-1,2j\}$ pair of sites.
Ultimately, the bulk, and thus the thermodynamical limit of this phase, will be strongly sensitive to the boundary conditions.
As local order parameter to capture this phase
we adopt the staggered spatial average $\eta$ of the matter-gauge interaction. Precisely, since we always simulate an even number of sites $\len$, we set
$\eta = | \frac{2}{\len} \sum_{j=1}^{\len/2} \langle H^{\text{inter}}_{2j-1,2j} \rangle - \frac{2}{\len - 2} \sum_{j=1}^{\len/2 - 1} \langle H^{\text{inter}}_{2j,2j+1} \rangle |$,
with $H^{\text{inter}}_{j,j+1} = \sum_{a,a'}^{r,g,b} ( c^{\dagger}_{j,a} \, U_{j,j+1}^{a,a'} \, c_{j+1,a'} + h.c. ) $,
as dimer
order parameter signaling a spontaneous breaking of the translation by one site, and observe it converge to a finite value in the dimer phase (see Fig.~\ref{fig:orderpars}).
We located the critical values $t^{(A)}_{c}$ and $t^{(B)}_{c}$ via a standard finite-size scaling procedure \cite{FisherBarber}.
We also detected a narrow window $t^{(A)}_c < t < t^{(B)}_c$, at filling $\nu = 3/2$, where the competition between the two insulating orders actually
favors a liquid order: within this window we again observe the Luttinger liquid order of Baryons, identified by $\xi_3$ (also shown in Fig.~\ref{fig:orderpars}).
We interpret this results as a vanishing
of the mass gap between the lower and the upper bands of the Baryon conductor, which occurs only within this small window of the $t/g$ ratio.
For any filling different from the KS vacuum, that is $\nu \neq 3/2$, the system behaves as a band conductor of Baryons.
Specifically, we observe a divergence with $\len$ of the correlation length $\xi_3$ related to the Luttinger liquid order of Baryons. Similarly,
we fit a nonzero central charge $c \geq 1$ (roughly $c \simeq 1.3 \pm 0.1$) and a Fermi wavevector $k_F \simeq \frac{\pi}{3} \nu$,
suggesting that we have a single band of weakly interacting, quasi-free baryons for $0 < \nu < 3/2$ and another band for $3/2 < \nu < 3$.
When considering finite bare masses $m > 0$ we detect two main changes in the phase diagram. First of all the narrow window of Baryon liquid at
the KS vacuum filling $\nu = 3/2$ disappears completely, and we observe a simple transition between the (induced) chiral insulator and the dimer phases.
Moreover we observe the emergence of effectively insulating phases when $t \ll m$ for any filling $\nu$. However, this regime is extremely hard to simulate via
imaginary TEBD and we can not rule out metastabilities.
 
{\it (3) Binding energies $-$}
We also studied the mass gaps, at finite size, of the SU(3) lattice gauge theory. This is possible thanks to our canonical treatment of the particle number symmetry, which allows us to study the system with one or a few excess quarks on top of the KS vacuum, and thus measure the binding energies between them. This analysis revealed that while each baryon (three quarks) is a strongly bound state, two baryons (six quarks) weakly repel each other. This suggest that (flavorless) lattice QCD in 1D disfavors the creation of multi-baryon bound states, such as atomic nuclei, in sharp contrast with QCD
in three dimensions (see SM for more details).

{\it Conclusions $-$} 
We simulated the equilibrium properties of a compact one-dimensional flavorless lattice QCD: the theory is strongly confined, i.e., it exhibits only colorless quasiparticles.
Our approach shows the power of Tensor Network for treating non-perturbatively lattice gauge theories in low dimensions, while exactly capturing the gauge symmetry content, even when it is a complex non-Abelian structure such as the SU(3) group.

\begin{acknowledgments}
We warmly thank A. Celi, K. Jansen and P. Zoller
for stimulating private discussions.
Authors kindly acknowledge financial support
from the EU via ERC Synergy Grant UQUAM, the QuantERA project QTFLAG, and the Quantum Flagship PASQUANS, 
 from the Baden-W\"urttemberg Stiftung via Eliteprogramm for PostDocs, from the Carl-Zeiss-Stiftung via Nachwuchsf\"orderprogramm, 
  and from the DFG via the TWITTER project.
\end{acknowledgments}

%

\newpage

\section{Supplementary Material}
\appendix

\section{A1. Gauss' Law}

The SU(3) Gauss' law at every site thus reads $Q^{\nu}_{j} |\Psi_{\text{phys}} \rangle = 0$, $\forall \nu \in \{1,8\}$ and $\forall j \in \{1,\len\}$, where
\begin{equation} \label{eq:gauss}
 Q^{\nu}_{j} = R^{\nu}_{j-1,j} + \sum_{a,a'}^{r,g,b} \frac{\lambda_{a,a'}^\nu}{2} c^{\dagger}_{j,a} c_{j,a'} \ +  L^{\nu}_{j,j+1},
\end{equation}
unambiguously determining the subspace of physical states $|\Psi_{\text{phys}} \rangle$.
Equivalently, the Gauss' law is requiring that each matter field, fused with the gluon to its left (right-hand side)
and with the gluon to its right (left-hand side) forms a color singlet $(0,0)$ overall.

\section{A2. Quantum Link Model derivation}

In this section we discuss how to explicitly obtain the 19-states truncated $SU(3)$ Lattice Gauge model under our study from a Quantum Link Model perspective \cite{ChandraJens,Brower1999}.
This pathway allows us to easily compute all the relevant matrix elements for the various effective operators of the many-body model, so that we can export
them on a numerical simulation platform (see next section).
In this paradigm, each gauge field degree of freedom $\{j,j+1\}$ is replaced with a pair of quark-colored fermionic modes, or `rishons', $\{j,L\}$ and $\{j+1,R\}$. Hence, the substitutions
\begin{equation}
\begin{aligned}
U_{j,j+1}^{a,a'} &\to \psi_{j,L,a} \psi^{\dagger}_{j+1,R,a'} \;, \\
L^{\nu}_{j,j+1} &\to \frac{1}{2} \sum_{a,a'} \psi^{\dagger}_{j,L,a} \lambda^{\nu}_{a,a'} \psi_{j,L,a'} \quad \mbox{and}\\
R^{\nu}_{j,j+1} &\to \frac{1}{2} \sum_{a,a'} \psi^{\dagger}_{j+1,R,a} \lambda^{\nu}_{a,a'} \psi_{j+1,R,a'},
\end{aligned}
\end{equation}
provide the correct commutation relations between $U_{j,j+1}^{a,a'}$ and $L^{\nu}_{j,j+1}$ (resp.~$R^{\nu}_{j,j+1}$). Such splitting generates an Abelian link-symmetry as a byproduct,
corresponding to $\mathcal{N}_{j,j+1} = n_{j,L} + n_{j+1,R} = \sum_{a} \psi^{\dagger}_{j,L,a} \psi_{j,L,a} + \sum_a \psi^{\dagger}_{j+1,R,a} \psi^{\dagger}_{j+1,R,a}$. Specifically, we choose to work
in the symmetry sector $\mathcal{N}_{j,j+1} |\Psi_{\text{phys}}\rangle = 3 |\Psi_{\text{phys}}\rangle$: This produces automatically a 20-dimensional gluon field space which
contains the desired 19-states truncated space. Precisely, the two states with $n_{j,L} = 0$ and $n_{j,L} = 3$ transform both as colorless-colorless $(0,0) \otimes (0,0)$ states,
while the nine states with $n_{j,L} = 1$ correspond to the quark-antiquark $q \otimes \bar{q}$ states, and finally the nine states $n_{j,L} = 2$
correspond to the antiquark-quark states $\bar{q} \otimes q$.
Under these considerations, we can to formally rewrite the (bare) free-field term of the truncated lattice QCD Hamiltonian as
\begin{equation}
 H_{\text{ff}} = \frac{g_0}{4} \sum_{j=1}^{\len -1} n_{j,L} (3 - n_{j,L}) + n_{j+1,R} (3-n_{j+1,R}),
\end{equation}
where now the two colorless-colorless states have energy $0$, and the other 18 states have energy $g_0$.

In addition to the aforementioned splitting, Quantum Link formulations of gauge theories add an extra term in the Hamiltonian whose purpose is to break the artificial symmetry
$\mathcal{Q}_j = n_{j,R} + \sum_{a} c^{\dagger}_{j,a} c_{j,a} + n_{j,L}$. This symmetry is unwanted, as it changes the gauge transformations from an SU(3) group into a U(3) group.
The dynamical breaking terms thus typically takes the form:
\begin{multline}
 H_{\text{break}} = - \varepsilon \sum_{j=1}^{\len -1} \left( \psi^{\dagger}_{j,L,r} \psi^{\dagger}_{j,L,g} \psi^{\dagger}_{j,L,b} \times \right. \\ \left. \times \psi_{j+1,R,b}  \psi_{j+1,R,g}  \psi_{j+1,R,r} + h.c. \right),
\end{multline}
also acting solely on the gauge degrees of freedom and
rightfully preserving all the other symmetries. This extra component has also the merit of breaking the energy degeneracy of the two colorless-colorless gluon states,
producing one reflection-symmetric state at energy $- \varepsilon$ and one reflection-antisymmetric state at energy $+ \varepsilon$.
We can now recover exactly the desired 19-states representation by energetically eliminating one of these two states, {\it i.e.}~by setting
$g_0 = (\frac{8}{3} g \ell - \varepsilon)$ and $\varepsilon \gg g \ell $. As we approach $\varepsilon \to +\infty$, the original free-field hamiltonian is recovered (apart from a constant).


\section{A3. Composite-site gauge invariant basis and gauge group reduction}

The following sections we perform some analytic manipulations of the truncated lattice QCD model in order to simplify the many-body quantum
problem in a format which is fit for numerical simulation. First of all we construct all the possible gauge-invariant quasi-local states, and show that it is possible to express them
in a color-transparent occupation basis, on which the gauge symmetry can be cast as a $Z_3$ group. We will then be able to evaluate all the matrix elements
of the truncated lattice QCD Hamiltonian on this basis.

To represent the matter states, we write sites $|0\rangle$ as the bare staggered vacuum (empty matter, full antimatter),
then the singly-occupied states $|r\rangle = c^{\dagger}_r|0\rangle$,
$|g\rangle = c^{\dagger}_g|0\rangle$ and $|b\rangle = c^{\dagger}_b|0\rangle$ (red, green, and blue respectively),
followed by the doubly-occupied states $|y\rangle = c^{\dagger}_r c^{\dagger}_g |0\rangle$,s
$|c\rangle = c^{\dagger}_g c^{\dagger}_b |0\rangle$ and
$|m\rangle = - c^{\dagger}_r c^{\dagger}_b |0\rangle$ (cyan, yellow and magenta),
and finally the triply-occupied state $|3\rangle = c^{\dagger}_r c^{\dagger}_g c^{\dagger}_b |0\rangle$.

To represent the rishon sites (either left or right side of a gluon field) we similarly adopt
$|r\rangle$, $|g\rangle$, $|b\rangle$
to characterize the quark-like $q$ subspace, 
$|y\rangle$, $|c\rangle$, $|m\rangle$
to characterize the antiquark-like $\bar{q}$ subspace, 
and $|0\rangle$ to represent the colorless $(0,0)$ subspace.

We can now list the gauge-invariant states of the quasi-local space, which includes a matter site $j$, the right-rishon space $R$ of the gluon to its left $\{j-1,j\}$
and the left-rishon space $L$ of the gluon to its right $\{j,j+1\}$:
\begin{equation}
\begin{aligned}
 &|0,0,0\rangle \\
 &|1,0,2\rangle &:=& \frac{1}{\sqrt{3}} \left( |r,0,c\rangle + |g,0,m\rangle + |b,0,y\rangle \right) \\
 &|2,0,1\rangle &:=& \frac{1}{\sqrt{3}} \left( |c,0,r\rangle + |m,0,g\rangle + |y,0,b\rangle \right)  \\
 &|0,1,2\rangle &:=& \frac{1}{\sqrt{3}} \left( |0,r,c\rangle + |0,g,m\rangle + |0,b,y\rangle \right)  \\
 &|1,1,1\rangle &:=& \frac{1}{\sqrt{6}} \left( |r,g,b\rangle + |g,b,r\rangle + |b,r,g\rangle + \right.\\ 
 &&& \left. - |g,r,b\rangle - |r,b,g\rangle - |b,r,g\rangle \right) \\
 &|2,1,0\rangle &:=& \frac{1}{\sqrt{3}} \left( |c,r,0\rangle + |m,g,0\rangle + |y,b,0\rangle \right) \\
 &|1,2,0\rangle &:=& \frac{1}{\sqrt{3}} \left( |r,c,0\rangle + |g,m,0\rangle + |b,y,0\rangle \right) \\
 &|2,2,2\rangle &:=& \frac{1}{\sqrt{6}} \left( |c,y,m\rangle + |y,m,c\rangle + |m,c,y\rangle + \right.\\ 
 &&& \left. - |c,m,y\rangle - |m,y,c\rangle - |y,c,m\rangle \right) \\
 &|0,3,0\rangle \\
 &|1,3,2\rangle &:=& \frac{1}{\sqrt{3}} \left( |r,3,c\rangle + |g,3,m\rangle + |b,3,y\rangle \right) \\
 &|2,3,1\rangle &:=& \frac{1}{\sqrt{3}} \left( |c,3,r\rangle + |m,3,g\rangle + |y,3,b\rangle \right),
\end{aligned}
\end{equation}
for a total of 12 composite-site gauge invariant states.
It is immediately visible that such occupation basis expansion $|\eta_R,\eta_M,\eta_L\rangle$, with $\eta_R, \eta_L \in \{0,1,2\}$ and $\eta_M \in \{ 0,1,2,3\}$,
bears no further color ambiguity once the Gauss' law is set, so it can be comfortably used for numerical simulations.

Additionally, in this basis it is straightforward to see that the gluon space truncation we chose has the net effect of simplifying the $SU(3)$ gauge symmetry
into its center group $\mathbb{Z}_3$, the Abelian cyclic group with three elements ($g$, $g^2 = g^{-1}$ and $g^3 = 1$). Namely,
let us define at every composite site $j$ the unitary operator $\Gamma |\eta_R,\eta_M,\eta_L\rangle = e^{\frac{2 i \pi}{3} (\eta_{R} + \eta_{M} + \eta_{L})} |\eta_R,\eta_M,\eta_L\rangle$.
The transformation $\Gamma$ clearly forms a $\mathbb{Z}_3$ group since $\Gamma^3 = \Id$.
In the truncated space, the Gauss' law of Eq.~\eqref{eq:gauss} can be simply cast in terms of the $\Gamma$ operators and reads
$ \Gamma_j |\Psi_{\text{phys}}\rangle = |\Psi_{\text{phys}}\rangle$, $\forall j \in \{1,\len\}$.
In simple terms, the reduced Gauss' law requires that, at every site $j$, $\eta_{R} + \eta_{M} + \eta_{L}$ modulo $3$ is equal to zero, which is indeed
a $\mathbb{Z}_3$ symmetry constraint.
As a final remark, we also point out that the link symmetry constraint can be similarly cast in a $\mathbb{Z}_3$ fashion,
since it can be written as $(\eta_{j,L} + \eta_{j+1,R}) \% 3 = 0$

\section{A4. Matrix Elements of the Hamiltonian}

In this section we derive the matrix elements for the operators that appear in the hamiltonian \eqref{eq:hamzero}, when expressed in the composite-site gauge
invariant basis we constructed in Sec.~A2. These effective operators directly follow from the quantum link modelization (see Sec.~A1), from which we projected away
the reflection-antisymmetric colorless-colorless state
\begin{multline}
|\infty \rangle_{j,j+1} = \frac{1}{\sqrt{2}} (\psi^{\dagger}_{j,L,r} \psi^{\dagger}_{j,L,g} \psi^{\dagger}_{j,L,b} + \\ + \psi^{\dagger}_{j+1,R,r} \psi^{\dagger}_{j+1,R,g} \psi^{\dagger}_{j+1,R,b}) | \Omega \rangle
\end{multline}
where $| \Omega \rangle$ is the particle vacuum (full antimatter universe in KS language). The effective 19-gluon-states Hamiltonian is then obtained from
the 20-gluon-states one in a first-order perturbation theory sense
$H^{(1)} = P H P$ where the projector $P = \bigotimes_j^{\len - 1} (\Id - |\infty \rangle \langle \infty |)_{j,j+1}$ removes the unwanted colorless-colorless state from each link.

According to this prescription, the various components of the truncated Hamiltonian \eqref{eq:hamzero} can be expressed as simple operators in the color-transparent occupation basis:
The free field term reads
\begin{equation} \label{eq:effectiveff}
 H_{\text{ff}}^{(1)} = \frac{2g \ell}{3} \sum_{j=1}^{\len -1} \eta_{j,L} (3 - \eta_{j,L}) + \eta_{j+1,R} (3-\eta_{j+1,R}) \; ,
\end{equation}
while the bare mass term reads
\begin{equation}
 H_{\text{bm}}^{(1)} = m \sum_{j=1}^{\len -1} (-1)^{j} \; \eta_{j,M} \; ,
\end{equation}
where now $\eta$ are meant as operators. The matter-gauge interaction term of the Hamiltonian can be comfortably written as
$H_{\text{inter}}^{(1)} = - \frac{t}{\ell} \sum_{j=1}^{L-1} A^{L \dagger}_j A^{R}_{j+1} + h.c.$ where the single composite-site operators $A^{L \dagger}$ and $A^{R}$ are
\begin{multline}
 A^{L \dagger} =  \sqrt[4]{\frac{9}{2}} |0,3,0\rangle \langle 0,2,1| + 2 |0,2,1\rangle \langle 0,1,2| \\
  +\sqrt[4]{\frac{9}{2}} |0,1,2\rangle \langle 0,0,0|  + \sqrt[4]{2} |1,2,0\rangle \langle 1,1,1| \\
  + \sqrt{2} |1,1,1\rangle \langle 1,0,2| + \sqrt[4]{\frac{1}{2}} |1,3,2\rangle \langle 1,2,0| \\
 + \sqrt{2} |2,3,1\rangle \langle 2,2,2| + \sqrt[4]{2} |2,2,2\rangle \langle 2,1,0| \\
 + \sqrt[4]{\frac{1}{2}} |2,1,0\rangle \langle 2,0,1|\,,
\end{multline}
and
\begin{multline}
 A^{R} = \sqrt[4]{\frac{9}{2}} | 0,0,0 \rangle \langle 2, 1,0| + 2 | 2, 1,0\rangle \langle 1, 2,0| \\
 + \sqrt[4]{\frac{9}{2}} |1,2,0\rangle \langle 0,3,0 | + \sqrt{2} |2,0,1\rangle \langle 1,1,1| \\
 + \sqrt[4]{2} |1,1,1\rangle \langle 0,2,1| + \sqrt[4]{\frac{1}{2}} |0,2,1\rangle \langle 2,3,1| \\
 + \sqrt[4]{2} |0,1,2\rangle \langle 2,2,2| + \sqrt{2} |2,2,2\rangle \langle 1,3,2| \\
 + \sqrt[4]{\frac{1}{2}} |1,0,2\rangle \langle 0,1,2|
\end{multline}
respectively. These matrices can be comfortably used in numerical simulations, such as the imaginary TEBD we employed to obtain our results.

\section{A5. String operators for fluid correlation functions}

In this paragraph we construct explicitly the string operator $S_{j,j'}^{a,a'}$ that we use in the evaluation of single-quark fluidity correlation matrix $\langle \mathcal{C}_{j,j'} \rangle$.
As mentioned in the letter, we require that it is a unitary space propagator (in the truncated model), and that it transforms covariantly, so that
the correlator $\mathcal{C}_{j,j'} = \sum_{a,a'} c^{\dagger}_{j,a} S^{a,a'}_{j,j'} c_{j',a'}$ preserves the Gauss' law at every site.

For such construction, it is extremely convenient to work in the occupation basis $|\eta_R,\eta_M,\eta_L\rangle$, on which the Gauss law is expressed as the
$\mathbb{Z}_3$ symmetry $[(\eta_{j,R} + \eta_{j,M} + \eta_{j,L}) \% 3 ] |\Psi_{\text{phys}}\rangle= 0$, and the link symmetry as
$[(\eta_{j,L} + \eta_{j+1,R}) \% 3 ] |\Psi_{\text{phys}}\rangle= 0$.

In this language, the correlators are simply expressed as
\begin{equation}
 \mathcal{C}_{j,j'} = T^{M \dagger}_{j} \left( \prod_{k=j}^{j'-1} B_{k,k+1} \right) T^{M}_{j'},
\end{equation}
where the order of the operators is not relevant as they all commute. In fact, the `tail' operators $T^{M}$ act only on matter modes, and retain some freedom: for simplicity, we chose them
to be
\begin{equation}
 T_{M} = |0\rangle \langle 1| + |1\rangle \langle 2| + |2\rangle \langle 3|.
\end{equation}
Each `body' operator $B_{j,j+1}$, instead, acts only on rishon modes $\{j,L \}$ and $\{j+1,R \}$. In order to preserve all link and Gauss' symmetries, they must be of the following form (apart from phase factors)
\begin{equation}
 B_{j,j+1} = |0,0\rangle \langle 1,2| + |1,2 \rangle \langle 2,1| + |2,1 \rangle \langle 0,0|,
\end{equation}
which is a cyclic operator $B^3 = \Id$.
We employ those string operators to investigate the presence of various Luttinger liquid orders: Respectively, liquid phases where the quasiparticles are quark-like ($\mathcal{C}_{j,j'} $),
meson-like ($\mathcal{C}^{2}_{j,j'} $) or baryon-like ($\mathcal{C}_{j,j'}^3 $).

\section{A6. Perturbation theory at $t \ll g$}

In this section we will derive an effective Hamiltonian, corresponding to the second-order degenerate perturbation theory in $t$, acting upon the ground space
of the free-field Hamiltonian, to better understand the phase diagram in the limit $t \ll g$ (after setting $\ell = 1$).
First of all we identigy the ground space of the free-field Hamiltonian: Using the free-field term in the form of Eq.~\eqref{eq:effectiveff}, we immediately
see that every link $\{j,j+1\}$ must be in the colorless-colorless state $|0,0\rangle$, kernel of the positive Casimir operators $C^{(L)}_j$ and $C^{(R)}_{j+1}$.
Due to the reduced $\mathbb{Z}_3$ Gauss' law, and having set the leftmost boundary to be colorless, we conclude that each matter site can only be
either in an empty state or in a triply-occupied (baryon) state, thus either $|0\rangle$ or $|3\rangle = c^{\dagger}_r c^{\dagger}_g c^{\dagger}_b |0\rangle$. The ground space of the free-field Hamiltonian is thus
equivalent to a 1D lattice of spinless fermions, given that baryon operators $b^{\dagger} = c^{\dagger}_r c^{\dagger}_g c^{\dagger}_b$ mutually anticommute as Dirac fermions.

First order perturbation theory on this effective subspace provides no change, since every matrix element of the matter-field interaction Hamiltonian restricted to this subspace is identically zero:
$P H^{(1)}_{\text{inter}} P = 0$ where $P$ is the projector onto the subspace.
At this point we can consider second-order processes on each pair of neighbouring sites, since the free-field Hamiltonian is fully local and the matter-field interaction acts on nearest-neighbours \cite{SWtrafo}.
States of the type $|0,0,0\rangle_j |0,0,0\rangle_{j+1}$ and $|0,3,0\rangle_j |0,3,0\rangle_{j+1}$ allow for no second-order processes.
On the other hand, the state $|0,0,0\rangle_j |0,3,0\rangle_{j+1}$  can virtually excite to the state $|0,1,2\rangle_j |1,2,0\rangle_{j+1}$ via $H^{(1)}_{\text{inter}}$, with amplitude $- \frac{3 \sqrt{2}}{2} t$.
In turn, the state $|0,1,2\rangle_j |1,2,0\rangle_{j+1}$, after freely propagating with inverse excitation energy $(H^{(1)}_{\text{ff}} - E_0)^{-1}$ equal to $\frac{3}{8g}$, can only go back
to the subspace in a single $H^{(1)}_{\text{inter}}$ excitation by returning to $|0,0,0\rangle_j |0,3,0\rangle_{j+1}$. Using the previous argument, we conclude that the state
$|0,0,0\rangle_j |0,3,0\rangle_{j+1}$ acquires a perturbed energy equal to $- 27 t^2/(16 g)$. By left-right reflection symmetry, the state
$|0,3,0\rangle_j |0,0,0\rangle_{j+1}$ acquires the same energy. The effective Hamiltonian
$H^{(2)} = P H^{(1)}_{\text{inter}} (H^{(1)}_{\text{ff}} - E_0)^{-1} H^{(1)}_{\text{inter}} P$
deriving from the second-order pertutbation theory in $t$ will thus read
\begin{equation}
  H^{(2)} = + \frac{27 t^2}{32 g} \sum_{j}^{\len-1} \sigma^{z}_j \sigma^{z}_{j+1}
  - \frac{3m}{2} \sum_{j=1}^{\len} (-1)^{j} \sigma^{z}
\end{equation}
where we re-introduced the bare mass term (which is exact, since it commutes with the free-field Hamiltonian), and where we recast the baryon lattice into
a spin lattice via Jordan-Wigner transformation, simply using $\sigma^{z}_j = 2 b^{\dagger}_j b_j -1$. The additional parameter that we control in our simulations, the particle 
filling $\nu$, is equivalent to $\nu = \frac{3}{2} + \frac{1}{L} \sum_j \sigma^{z}_j$ in this language.

In the bare-massless case, such ZZ Hamiltonian is gapped at the filling sector $\nu = 3/2$ (corresponding to zero magnetization),
and it has a doubly-degenerate ground state, corresponding to the two Z-aligned antiferromagnets.
This degeneracy spontaneously breaks the full spin-flip (particle-hole) symmetry, and establishes the chiral order parameter $\zeta_{\pi}$, which indeed we observed
numerically for $t \leq t_c^{(A)} g$.

\begin{figure}
\includegraphics[width=1.0\columnwidth]{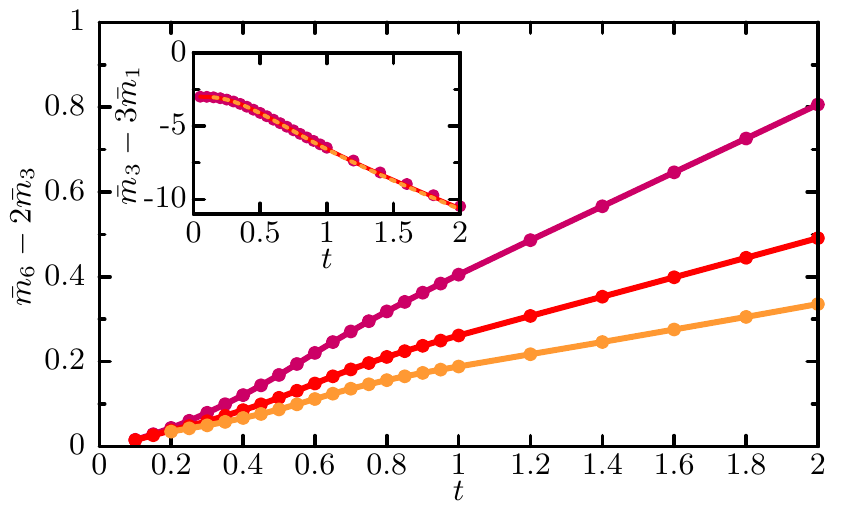}
\caption{ \label{fig:repel}
Binding energy $\bar{m}_6 - 2 \bar{m}_3$ of two baryons, as a function of $t$, for system sizes $\len = 48$ (magenta), $72$ (red), $96$ (orange). 
Inset: Binding energy  $\bar{m}_3 - 3 \bar{m}_1$ of the three quarks forming a single baryon. The sizes $\len$ are color coded as in the main panel.
}
\end{figure}

For other particle fillings $\nu$, at zero bare mass $m$, the effective second order Hamiltonian $H^{(2)}$ has an extensive degeneracy of the ground space, thus
to better understand the phases we find at $t \ll g$, it is helpful to consider additionally the third order of the perturbation theory:
Since $P H^{(1)}_{\text{inter}} P = 0$, the only relevant contribution to the third order effective Hamiltonian is of the form
$H^{(3)} = P H^{(1)}_{\text{inter}} (H^{(1)}_{\text{ff}} - E_0)^{-1} H^{(1)}_{\text{inter}} (H^{(1)}_{\text{ff}} - E_0)^{-1} H^{(1)}_{\text{inter}} P$,
which leads to
\begin{equation}
\begin{aligned}
 H^{(3)} &= \frac{81 t^3}{32 g^2} \sum_j^{\len-1} b^{\dagger}_{j} b_{j+1} + h.c. \\ &= \frac{81 t^3}{32 g^2} \sum_j^{\len-1} \sigma^{+}_{j}  \sigma^{-}_{j+1} + \sigma^{-}_{j}  \sigma^{+}_{j+1}
\end{aligned}
\end{equation}
where again we used the Jordan-Wigner. This effective tight-binding Hamiltonian for fermions produces a finite bandwidth for the free-fermion modes,
which in turn will produce a fermionic band-conducting phase for fillings $\nu \neq 3/2$. Nevertheless, activating the bare mass $m \neq 0$ dampens the hopping processes,
resulting again in an emergent insulating behaviour.

\section{A7. Binding energies}

Here we report the estimation of binding energies of few-body, eventually bound, states, based on the measurement of the mass gaps $\bar{m}_k$ of $k$ quarks on top of the dressed vacuum $\nu = \frac{3}{2}$.
To do so,
we numerically simulate the ground state energies $E(\nu)$ for $\nu = \frac{3}{2} + k/\len$, with excess quarks number $k$ from the set $k \in \{0,1,3,6\}$. The corresponding $k$-quarks mass gap (that is, the chemical potential for $k$ additional quarks) is thus $\bar{m}_k = E(\frac{3}{2} + k/\len) - E(\frac{3}{2})$. We can then compare the various $\bar{m}_k$ to estimate the binding energies of composite particles.

We first consider the binding energy of a baryon, formed by three quarks. Its binding energy, $\bar{m}_3 - 3 \bar{m}_1$ is reported in
Fig.~\ref{fig:repel} (inset) as a function of $t$, for zero bare mass $m$. We observe not only that this binding energy is negative, but also that it is insensitive to the system size. This suggest that the three quarks indeed form a bound state, which has finite size in the relative coordinates with respect to the center of mass, once again corroborating the strong confinement of quarks \cite{Wilson74} in our model.

Secondly, we consider the binding energy of a deuteron, i.e. a state formed by two baryons, or six quarks, hence $k = 6$. The main panel of Fig.~\ref{fig:repel} shows that the deuteron binding energy $\bar{m}_6 - 2 \bar{m}_3$ is positive, revealing that the two baryons do not form a bound state, while instead they repel each other. Moreover, we observe that such repulsion energy decreases with the system size. We consider this a signature that the effective interaction energy scales with the distance of the two baryons: as the size $\len$ increases, they have more space to sit far apart, and the resulting interaction energy decreases. The repulsion between baryons is thus somewhat `weak' compared to the color confinement effects.

%
%

\end{document}